\documentclass[aps,prb,floatfix,twocolumn,showpacs,amsmath,amssymb,eqsecnum,superscriptaddress]{revtex4}
\newcommand{\be}{\begin{equation}}
\newcommand{\ee}{\end{equation}}
\newcommand{\bea}{\begin{eqnarray}}
\newcommand{\eea}{\end{eqnarray}}
\newcommand{\ben}{\begin{eqnarray*}}
\newcommand{\een}{\end{eqnarray*}}

\usepackage{graphicx,color}
\usepackage{dcolumn}
\usepackage{bm}
\usepackage{amsmath}
\usepackage{amssymb}
\usepackage{mathrsfs}
\usepackage{float}
\usepackage{ulem}
\usepackage{epsfig}
\usepackage{enumerate}
\usepackage{epstopdf}

\begin{document}

\title{Engineering relativistic effects in ferroelectric SnTe}

\author{E. Plekhanov}
\affiliation{Consiglio Nazionale delle Ricerche - CNR-SPIN, I-67100 L'Aquila, Italy}
\author{P. Barone}
\affiliation{Consiglio Nazionale delle Ricerche - CNR-SPIN, I-67100 L'Aquila, Italy}
\author{D. Di Sante}
\affiliation{Consiglio Nazionale delle Ricerche - CNR-SPIN, I-67100 L'Aquila, Italy}
\affiliation{Department of Physical and Chemical Sciences, University of L'Aquila, Via Vetoio, I-67100 L'Aquila, Italy}
\author{S. Picozzi}
\affiliation{Consiglio Nazionale delle Ricerche - CNR-SPIN, I-67100 L'Aquila, Italy}

%\ead{evgeny.plekhanov@spin.cnr.it}

\begin{abstract}
Spin-orbit coupling is increasingly seen as a rich source of novel phenomena, as shown by the
recent excitement around topological insulators and Rashba effects. We here show that the addition
of ferroelectric degrees of freedom to a semiconductor featuring topologically-non-trivial
properties, such as SnTe, merges the intriguing field of spin-orbit-driven physics with non-volatile
functionalities appealing for spintronics. By using a variety of modelling techniques, we show that
a strikingly rich sequence of phases can be induced in SnTe, when going from a room-temperature
cubic phase to a low-temperature ferroelectric structure, ranging from a topological crystalline
insulator to a time-reversal-invariant $Z_2$ topological insulator to a ``ferroelectric Rashba
semiconductor", exhibiting a huge electrically-controllable Rashba effect in the bulk band
structure.
\end{abstract}
\pacs{71.20.-b,73.20.-r,71.70.Ej}
\maketitle

%\section{Introduction}
%{\bf Introduction.}
Many of the materials science topics that recently drew large attention arise from the intriguing
physics based on spin-orbit coupling (SOC). For example, growing enthusiasm is devoted to the so
called `Ferro-Electric Rashba Semi-Conductors (FERSC)", a novel class of multifunctional materials
featuring GeTe as prototype\cite{gete,FERSC}. In FERSC, a (giant) Rashba effect (RE) is present in
the {\it bulk} band structure and is intimately linked to ferroelectricity: the spin-texture can be
fully reversed when switching the ferroelectric (FE) polarization. Indeed, in Ref.~\onlinecite{gete} some
of the present authors predicted, by means of first-principles simulations, that FE GeTe would show
a huge Rashba spin-splitting for the valence band maximum and an unprecedented Rashba parameter.
While our theoretical predictions for GeTe urgently call for an experimental confirmation, the
fascinating FERSC phenomenology opens exciting perspectives in electrically-controlled
semiconductor spintronics: for example, a modified Datta-Das spin transistor based on a FERSC
channel would allow the appealing integration of storage and logic functionalities.\cite{gete} 
%in the same device.\cite{gete} 

Moreover, since few years topological insulators (TI) \cite{TI} have been at the center of
increasing excitements, due to exotic properties, such as their surface states (SS) showing a Dirac
spectrum and being exceptionally robust against perturbations. TI possess a non-trivial $Z_2$
classification, arising from an odd number of SOC-induced band inversions in their bulk
band-structure. The SOC-related origin of the phenomenon makes the SS to show a single
spin-state for each momentum, paving the way to spin-polarized currents of interest in spintronics.
Recently, a novel class of TI has been introduced, namely topological crystalline insulators (TCI),
\cite{TCI} where metallic edge states are protected by point- or space-group symmetries, rather than
time-reversal (TR) symmetry as in usual $Z_2$ TI. The proof of the existence of TCI came from rocksalt
SnTe \cite{SnTe1,SnTe2} (and related alloys\cite{SnTe3,SnTe4}), where the $fcc$ lattice shows a
(110) mirror plane and the (bulk) $L$ symmetry-point features a SOC induced band-inversion between
anion and cation in the valence-band-maximum (VBM) and conduction-band-minimum (CBM), compared to
other standard semiconductors. Surfaces preserving the mirror operation, such as the (001), were
first predicted (and later experimentally confirmed via angle-resolved photoemission) to show an
even number of Dirac branches along the $\overline{\Gamma}$-$\overline{X}$ direction, corresponding
to an even number of band inversions (rather than an odd number, as in standard TI).

In this work, we focus on the coexistence and interplay between FERSC characteristics and
topological behaviour, starting from the observation that any material showing at the same time
ferroelectricity and a topologically non-trivial character is likely to show large SOC and small
band gap: it is therefore expected to satisfy also the conditions for a (possibly large) bulk RE. A
multifunctional material being at the same time a FERSC {\it and} a TCI would clearly constitute a
breakthrough in the field and we show here that this is the case of SnTe. The role of displacements
in TCI was only marginally discussed in Ref.\onlinecite{SnTe1} for SnTe, where a FE distortion
along the [111] direction was considered
%: upon
%displacement (
(consistently with crystalline rhombohedral deformation, occurring
%in SnTe 
at low temperatures\cite{FeSnTe}).
%, out of the four Dirac points present in the [001] surface
%electronic structure for the cubic lattice, two were predicted to remain gapless (being protected by
%the - still present - mirror symmetry), whereas in the other two an energy gap was anticipated to
%occur. Indeed,
While Hsieh et al.~\cite{SnTe1} performed helpful symmetry considerations, they remained at the
speculative level. % as no quantitative calculations were carried out. 
Rather, we here show that it is important to perform a quantitative analysis. In closer detail, by
means of ab-initio simulations, we consider the transition from cubic to FE SnTe and explore
topological properties along the path connecting the two extremal phases, by focusing on [111]
surfaces. Indeed, several exciting outcomes result: {\it i}) below the FE transition temperature,
SnTe turns out to be a {\it novel FERSC}; {\it ii}) the TCI phase for the cubic rocksalt structure
can be turned to a $Z_2$ TI under external manipulation and {\it iii}) the {\it coexistence} of {\it
FERSC} behaviour with either {\it TCI} or $Z_2$ topological insulating phases is shown to occur.

\begin{figure*}
	\includegraphics[width=0.95\textwidth,angle=270,scale=0.37]{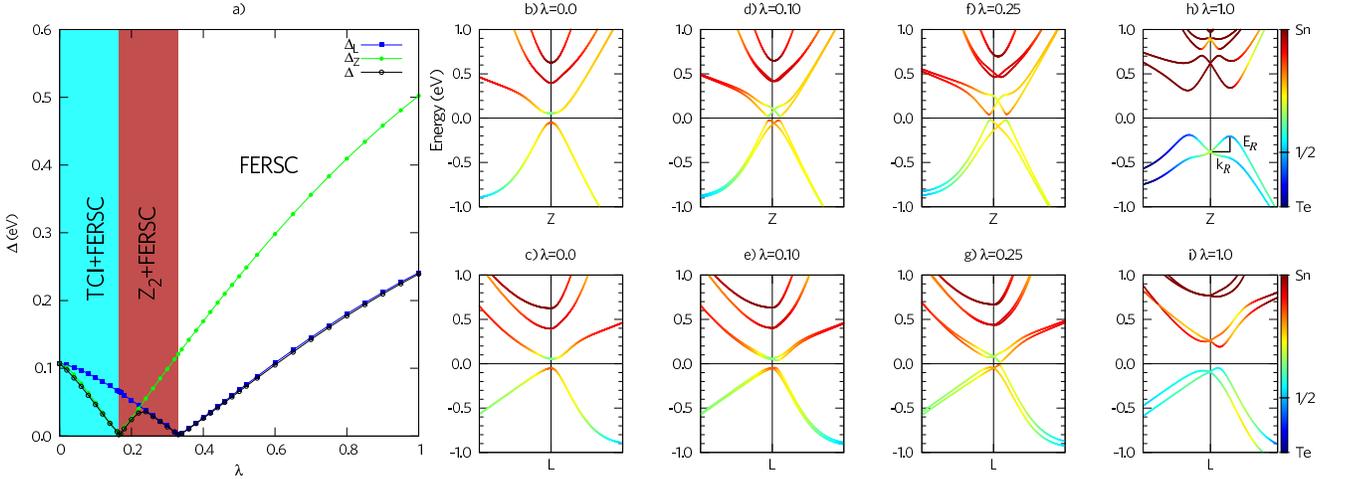} \caption{\label{fig1}
	   Ab-initio DFT: a) Topological phase-diagram estimated from the evolution, as a function of
	   $\lambda$, of the energy gaps around
	   %``in proximity" to 
	   $Z$ ($\Delta_Z$) and to $L$ ($\Delta_L$).
%       ,meaning that the gap is evaluated at the Rashba points $k_R$ around $Z$ and $L$,
%       respectively.
	   The ``minimal'' energy gap ($\Delta$) over the whole BZ is drawn in black. The
	   band structures in proximity to $Z$ (along the directions $B\to Z\to P$) and $L$ (along the
	   directions $\Gamma\to L\to B_1$), respectively, are shown in panels b) and c) for $\lambda$ =
	   0.0, in panels d) and e) for $\lambda$ = 0.10, in panels f) and g) for $\lambda$ = 0.25, in
	   panels h) and i) for $\lambda$ = 1.0. The orbital character of bands is shown by color map
	   going from dark blue (Te) to dark red (Sn), with $1/2$ corresponding to the equally mixed
	   anionic and cationic character. For clarity, we report in panel h) the graphical definition
	   of the Rashba momentum offset, $k_R$, and of the Rashba energy splitting, $E_R$.
	}
 \end{figure*}

{\it Technical details. }
Density functional theory (DFT) simulations were performed using the Vienna Ab initio Simulation
Package (VASP) \cite{vasp} and the Generalized Gradient Approximation (GGA)\cite{gga} in the
Perdew-Burke-Ernzerhof (PBE) formalism for the exchange-correlation potential. We used an energy
cutoff for the plane wave basis of $400$ eV and a $16\times 16\times 16$ Monkhorst-Pack $k-$point
mesh~\cite{mp}. Test calculations using more accurate hybrid functionals within the HSE\cite{hse}
formalism (see \cite{sm}) showed that the relevant physics was only marginally affected, so we will
focus here on the GGA results. We here concentrate on the [111] surface, usually a natural cleavage
plane for rhombohedral crystals. To calculate $(111)$ SS in the slab geometry, the
number of layers has to be sufficiently large to distinguish the surface from the bulk states
%; at critical regimes, when the energy gap is small, one needs to consider up to a thousand of layers, making 
which makes ab-initio approaches prohibitive. We therefore resorted to an effective tight-binding (TB) model
for SS. The TB hopping matrix elements were determined by projection of the
ab-initio VASP Hamiltonian onto the Maximally Localized Wannier Orbitals (MLWO) through the
WANNIER90 package~\cite{wannier90}. As for structural parameters, we employed those optimized
within DFT-GGA. For the rhombohedral (cubic) structure, we used lattice constants $a$ = 6.475 \AA
(6.420 \AA), rhombohedral angles $\alpha$ = 58.8$^o$ (60$^o$) and atomic displacements (in internal
units) $\tau = 0.026 (0)$ (see Section A and Supp. Tab. 1 of Ref.~\onlinecite{sm}.)

{\it Bulk electronic structure. }
As already mentioned, the cubic phase of SnTe is a TCI, while the rhombohedral phase is FE (with a
calculated polarization of $\sim40 \mu C/cm^2$, see Ref.~\onlinecite{sm}). We here search for an
intermediate phase (allowed by symmetry) which exhibits both ferroelectricity and RE, being at the
same time topologically non-trivial. In order to demonstrate the existence of such a phase, we
construct a path, parametrized by $\lambda$, linearly connecting the structural parameters (i.e. in
terms of lattice constants and angles, atomic positions, etc) of the cubic TCI phase (space group
$Fm3m$, $\lambda=0$) to the rhombohedral FE one (space group $R3m$, $\lambda=1$). 
%We remark that, in our
%simulations, $\lambda$ is a parameter quantifying the distortion from the cubic to the FE
%phase and related displacements, so that intermediate values of $\lambda$ might be experimentally
%difficult to achieve. However, our findings are far more general, as we expect that 
%While $\lambda$ represents here only a theoretical parameter,
%the predicted sequence of topologically interesting phases found below can be induced, controlled and
%optimized using 
%it can be realized by means of (external) physical handles, such as chemical doping or
%alloying, temperature, pressure, strain, etc.
%The results of o
Our DFT calculations %for geometries calculated within DFT-GGA 
are reported in Fig.~\ref{fig1}, where we show the energy gaps at the $Z$ and $L$ points of the
rhombohedral Brillouin zone (BZ), along with a zoom of the related band structures, for different
values of $\lambda$.

Let's first discuss the FERSC behaviour. As for GeTe,\cite{gete} SnTe meets all the necessary
conditions pointed out to support a giant RE. Specifically, the VBM and CBM, both at $Z$ and $L$
points, have the same symmetry character coming from an unusual bands' ordering close to the Fermi
level. This, in combination with a rather strong SOC, a very small gap and the lack of inversion
symmetry, paves the way for a huge electrically-controllable bulk Rashba
spin splitting~\cite{bahramy} (RSS): in the low-temperature phase ($\lambda = 1$), we calculated a Rashba
momentum offset, around $Z$, $k_R \approx$ 0.08 \AA$^{-1}$. The energy splitting at $Z$, $E_R$
[calculated as the energy difference between the lowest conduction (highest valence) band at $k_R$
and at the high symmetry points (HSP) $Z$] is as large as 272 (172) meV for the conduction (valence) bands,
see Fig.\ref{fig1} h). Being the Rashba parameter $\alpha_R$ as large as 6.8 (4.4) eV/\AA $\:$ for
the conduction (valence) bands, SnTe is unambiguously proven to be an additional example of FERSC
for all values $\lambda >$ 0 (see Section B. and Fig. Supp. 1 in Ref.~\onlinecite{sm} for further
details).
%on the FERSC properties
Moreover, differently from GeTe,\cite{gete} in SnTe the VBM and CBM at $Z$
both belong to the $j=1/2$ manifold (at variance with $j=1/2$ and $j=3/2$ for VBM and CBM,
respectively, in GeTe). Therefore, SnTe seems even more promising than GeTe, as the large RSS
occurs both in the valence and in the conduction bands, opening the routes towards an
ambipolar behaviour\cite{bitei} of interest for spintronics.% applications.
 
As for the topological properties, let's remark that, along the whole $\lambda$ path, SnTe is
characterized by the presence of a band gap among different Rashba peaks, $k_R$, around either $L$
or $Z$. The key point is, however, the band structure evolution, as a function of $\lambda$, of the
$Z$ and $L$ points, the two being different because of the rhombohedral/off-centering distortions.
By looking at the inverted orbital character (typical for TI), we note that the TCI phase survives
the rhombohedral distortion for quite long (up to $\lambda = 0.16$, cfr. Fig.\ref{fig1} a)). In
this range, ferroelectricity coexists with a small RSS, mainly located at $Z$; we
therefore label this phase as ``{\it TCI+FERSC}". For the critical value, $\lambda=0.16$, the gap
closes at $Z$. Upon larger displacements, the gap at $Z$ reopens without band inversion, whereas at
$L$ the band inversion is still present. In this range of $\lambda$, an extremely exotic phase
exists: along with the FERSC behaviour, an odd number of 
band inversions opens the possibility for a $Z_2$-type of TI. This phase, labeled as ``{\it
$Z_2$+FERSC}", is indeed confirmed to be a strong TI from the evaluation of the $Z_2$ topological
indices as $(1,111)$ (see Section C in Ref.~\onlinecite{sm} for details). Finally, a further increase in
$\lambda$ closes the gap (at $\lambda=0.33$) at the Rashba points $k_R$ in the vicinity of $L$, thus
switching the band inversion off (along with the TI regime) at $L$. The system consequently becomes
a trivial FE insulator with strong RE (i.e. a ``{\it topologically-trivial FERSC}"). The sequence
of phases (TCI+FERSC,$Z_2$+FERSC, FERSC), shown in Fig.\ref{fig1} a), is the main result of the
paper, further explored below in terms of surface electronic structure.

\begin{figure*}
	\includegraphics[width=0.95\textwidth,angle=270,scale=0.75]{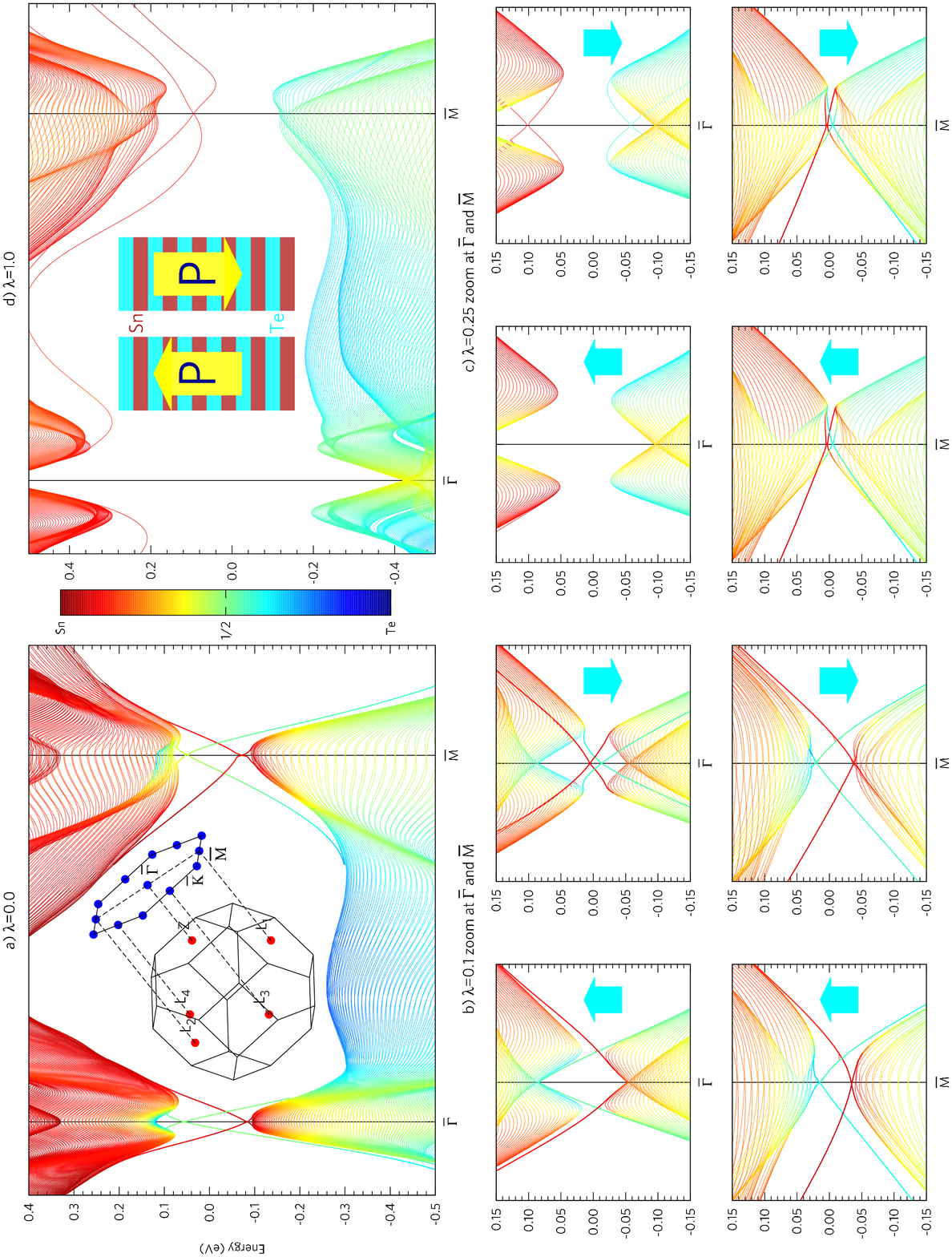} \caption{\label{fig2}
	   TB$+$MLWO: a) [111] SBS for $\lambda = 0.0$. The inset shows
	   the rhombohedral BZ and its projection on the [111] surface. d) [111] SBS
	   for $\lambda = 1.0$.  The inset shows the stacking sequence of the unit cells, along with the
	   two possible terminations and direction of polarization (labeled as $P_{\uparrow}$ in the left and
	   $P_{\downarrow}$ in the right). b) and c) show the SBS for $\lambda = 0.1$ and
	   $\lambda=0.25$. In the latter panels, there are four subpanels, showing the band structure
	   around $\overline{\Gamma}$ for $P_{\uparrow}$ (upper left), around $\overline{\Gamma}$ for
	   $P_{\downarrow}$ (upper right), around $\overline{M}$ for $P_{\uparrow}$ (lower left), around
	   $\overline{M}$ for $P_{\downarrow}$ (lower right). All the calculations were performed on slabs
	with $480$ Sn and Te layers, except for the case of $\lambda=0.25$ and around $\overline{M}$
 point (where higher accuracy is needed), in which $720$-layer slabs were used. Color scale as in
 Fig.1.}
\end{figure*}

{\it Surface states.}
Owing to the bulk-boundary correspondence (BBC), all the findings about topological states in the bulk
find their counterparts in the electronic properties of the [111] SS, showing
intriguing aspects related to ferroelectricity.
Since the FE polarization is perpendicular to the $(111)$ surface~\cite{unpub}, in the slab geometry with mixed
termination (Te on one side, Sn on the other) two possibilities can be distinguished: one, when
polarization points from the inner layers towards the Sn surface (and away from the Te surface) or
the other, when it points away from the Sn surface (and towards the Te surface), as shown in the
inset of Fig.~\ref{fig2} d). Due to the mixed termination, the contributions of each surface can be
distinguished by looking at their orbital and atomic character.
We underline that the presence or absence of the topological SS is indeed a
"bulk"-derived property, marginally depending on the surface band structure (SBS)
%. Furthermore, thanks to the BBC,
%the topological protection guarantees that SS survive even in 
or on the presence of surface defects. As a matter of fact, the topological nature of a band insulator is
routinely assessed by calculating SS in a slab geometry, often in the framework of
empirical TB models~\cite{SnTe4,superf_calcs}. Here, we use a much more realistic TB
parametrization derived from an accurate first-principles-based Wannier procedure.

As well known, in the SnTe cubic phase, each surface of the slab exhibits a cone of SS
at $\overline{\Gamma}$ and each of the three non-equivalent $\overline{M}$ points of the surface
Brillouin zone (cfr inset in Fig.~\ref{fig2} a), consistent with the existing
literature\cite{fu,buco} and with the band inversion observed in the bulk bands. It is worth noting
that Sn surface cones cross in the vicinity of the VBM, while the Te SS cross close to
the CBM.

In the TCI+FERSC phase (0$<\lambda <$0.16), the system shows a finite FE polarization but is
topologically equivalent to the Cubic one, {\it i.e.} we find in the surface electronic structure
the same number of cones at the same symmetry points for $\lambda$ = 0.0 and 0.1 [cfr Fig~\ref{fig2}
a) and b)]. However, for a finite $\lambda$, the behaviour at $\overline{\Gamma}$ depends on the
polarization direction: %since the Rashba peaks develop at $Z$ (projected to $\overline{\Gamma}$ in the slab
%geometry) and the SS have to cross at $\overline{\Gamma}$ as a TRIM point, 
the crossing can either occur close to the middle of the gap (cfr Fig.~\ref{fig2} b) upper-right
panel) or closer to crossing of projected bulk bands (cfr. Fig.~\ref{fig2} b) upper-left panel).
This allows for an additional control of the surface electronic properties through an electric
field.

In the $Z_2$+FERSC phase (0.16$<\lambda <$0.33), %a further rhombohedral distortion and FE
%off-centering induce the closure of the gap at $Z$ between the bands with 
the inverted (normal) gap at $L$ ($Z$) is reflected on the surface electronic structure [cfr. Fig.
~\ref{fig2} c)]. The striking feature of this phase is the presence of a {\it single Dirac cone}
from each surface crossing at $\overline{M}$. Since the number of nonequivalent Dirac cones is odd,
this phase is confirmed to be a $Z_2$ TR invariant strong TI with indices $(1,111)$. The
bulk bands (projected at the surface) exhibit a strong RE at $\overline{\Gamma}$ and a tiny one at
$\overline{M}$. Although the SBS at $\overline{M}$ does not depend on the FE
polarization direction, at $\overline{\Gamma}$ it does. Indeed, for polarization corresponding to
the negative shift of the Te sublattice with respect to the Sn one, a {\it surface RE} appears. We
recall, in fact, that the RE discussed so far in the paper occurs in the {\it bulk}; however, a
more ``conventional" RE can also occur at the {\it surface}, deriving from the breaking of inversion
symmetry induced by the presence of any surface. %The SS connecting the bulk Rashba peaks in VBM
%are mainly of Te character, while in CBM they are mainly to be ascribed to Sn.

Recall now that, in the bulk, %the closure of the gap at $L$ occurs at $\lambda=0.33$. % and is
%accompanied by the interchange of orbital character. T
%Therefore, 
the phase at $\lambda>0.33$ does not show any band inversion and is topologically trivial.
This phase also characterizes the final end-point of our adiabatic
path at $\lambda=1$ (FERSC phase), for which we show the SBS in Fig.~\ref{fig2}
d). Here, the bulk RE is evident both at $\overline{\Gamma}$ and $\overline{M}$, while the surface
RE is present on one of the two HSP depending on the orientation of polarization.

In summary, we have addressed the intriguing physics brought by SOC interlinked with
ferroelectricity in SnTe. By means of a combination of DFT, TB and
Wannier orbitals, we analyzed bulk and surface electronic structures and showed that, upon any
external agent (such as pressure, strain or chemical doping) that is able to drive a transition from
the room-temperature cubic centrosymmetric structure to the low-temperature FE structure of SnTe, an
exotic sequence of peculiar phases can be induced, controlled and optimized. Indeed, these phases
range from a TCI for small deformations ($0< \lambda< 0.16$) to a
$Z_2$ topological insulator for slightly larger FE distortions ($0.16< \lambda< 0.33$) to a trivial
topological behaviour up to the experimental FE structure ($0.33< \lambda< 1$). Moreover, we show
that topological phases can coexist with a FERSC behaviour, where a strong RSS
is predicted to occur in the bulk electronic structure and whose spin texture is
controllable via an electric field.
While the specific range of $\lambda$ values for each phase depends on model
details, we predict the {\it sequence of phases} - as a function of $\lambda$ - from TCI over $Z_2$
to trivial insulator with strong RE, to be quite general.
%Furthermore, we comment on
%For what regards
the possibility to experimentally observe 
As for the experimental realization of the intermediate phases, we note that the
%(TCI$+$FERSC, $Z_2+$FERSC, FERSC). The
parameter $\lambda$, connecting the cubic and the
%off-centered 
rhombohedral phases, is mainly
related to the ferroelectric displacement between Sn and Te sublattices and, hence, can be controlled by
several means, such as $p$-doping or chemical substitution. Indeed, the naturally occurring
$p$-type doping in SnTe, due to Sn vacancies, strongly affects FE properties. For example,
$p$-doping above a critical threshold~\cite{FeSnTe} 
%($\approx 1.5\times 10^{20} cm^{-3}$)
completely destroys the
displacement. In addition, it has been theoretically argued that $n$-type doping could smoothly tune
the atomic displacement in the prototypical FE, BaTiO$_3$~\cite{yongwang}.
%We, therefore, expect the tuning of the $p$-type doping to act in a similar way in SnTe.
Analogously,
the study of isovalent substitution in SnGeTe~\cite{gesnte} showed that the displacement can be
controlled by tuning the concentration of Sn.
While our results urgently call for an experimental verification, it is clear
that the multifunctional behaviour of FE SnTe shows a wide tunability through a variety of different
topological insulating or FERSC
phases.
More generally, our work shows how the peculiar interplay among ferroelectricity,
SOC, topology and RE opens exciting perspectives in different key areas
in current science, ranging from fundamental condensed matter physics (in terms of microscopic
mechanisms), to materials science (in terms of novel and advanced compounds) to technology (in terms
of a new generation of electrically-controlled spintronic devices).

%{\bf Acknowledgements} 
We acknowledge computational support through the PRACE project "TRASFER", CINECA facilities available through the ISCRA initiative (FeBTOMEC project) and the Supercomputing Cluster at CNR-SPIN SA (CLUSA).

%\bigskip 


\begin{thebibliography}{100}
\bibitem{gete} D. Di Sante, P. Barone, R. Bertacco and S. Picozzi, 
%Electric Control of the Giant Rashba Effect in Bulk GeTe, 
Adv. Mater. {\bf 25}
, 509 (2013); ibid. {\bf 25}, 3620Ð3626 (2013).

\bibitem{FERSC} S. Picozzi, Front. Physics, {\bf 2}, 10 (2014). %Ferroelectric Rashba Semiconductors as a novel class of multifunctional materials, 

\bibitem{TI} See for example M. Z. Hasan, C. L. Kane, Topological insulators, Rev. Mod. Phys. 82, 3045 (2010); X.L. Qi and S. C. Zhang, %Topological insulators and superconductors, 
Rev. Mod. Phys. 83, 1057 (2011); J.E. Moore, %The birth of topological insulators, 
Nature 464, 194 (2010); L.Fu, C.L. Kane and E. J. Mele, %Topological Insulators in Three dimensions, 
Phys. Rev. Lett. 98, 206803 (2007).

\bibitem{TCI} L. Fu, %Topological Crystalline Insulators, 
Phys. Rev. Lett. 106, 106802 (2011)

\bibitem{SnTe1} T. H. Hsieh, H. Lin, J. Liu, W. Duan, A. Bansil, and L. Fu, %Topological crystalline insulators in the SnTe material class, 
Nat. Comm. 3, 982 (2012).

\bibitem{SnTe2} Y. Tanaka, Zhi Ren,	T. Sato, K. Nakayama, S. Souma,	T. Takahashi, K. Segawa, and Y. Ando, %Experimental realization of a topological crystalline insulator in SnTe, 
Nature Phys. 8, 800 (2012).

\bibitem{SnTe3} S.Y. Xu, C. Liu, N. Alidoust, M. Neupane, D. Qian, I. Belopolski, J.D. Denlinger, Y.J. Wang, H. Lin, L.A. Wray, G. Landolt, B. Slomski, J.H. Dil, A. Marcinkova, E. Morosan, Q. Gibson, R. Sankar, F.C. Chou, R.J. Cava, A. Bansil, M.Z. Hasan, %Observation of a topological crystalline insulator phase and topological phase transition in Pb1? xSnxTe, 
 Nature Comms. 3, 1192 (2012).
 
\bibitem{SnTe4} P. Dziawa, B. J. Kowalski, K. Dybko, R. Buczko,	A. Szczerbakow,	M. Szot, E. {\L}usakowska, T. Balasubramanian, B. M. Wojek, M. H. Berntsen, O. Tjernberg, and T. Story, %Topological crystalline insulator states in Pb$_{1-x}$Sn$_{x}$Se, 
Nature Mater. 11, 1023 (2012).

\bibitem{FeSnTe} M. Iizumi, Y. Hamaguchi, K. F. Komatsubara and Y. Kato, %Phase Transition in SnTe
%with Low Carrier Concentration, 
J. Phys. Soc. Jpn. 38, 443 (1975). 

\bibitem{vasp} G. Kresse and D. Joubert, %From ultrasoft pseudopotentials to the projector augmented-wave method., 
Phys. Rev. B {\bf 59}, 1758-1775 (1999).

\bibitem{gga} J. P. Perdew, K. Burke and M. Ernzerhof, %Generalized Gradient Approximation Made Simple. 
Phys. Rev. Lett. {\bf 77}, 3865-3868 (1996).

\bibitem{bitei} A. Crepaldi, L. Moreschini, G. Autès, C. Tournier-Colletta, S. Moser, N. Virk, H. Berger, Ph. Bugnon, Y. J. Chang, K. Kern, A. Bostwick, E. Rotenberg, O. V. Yazyev, and M. Grioni, %Giant Ambipolar Rashba Effect in the Semiconductor BiTeI. 
Phys. Rev. Lett. 109, 096803 (2012).

\bibitem{wannier90} A. Mostofi, J. R. Yates, Y.-S. Lee, I. Souza, D. Vanderbilt, and N. Marzari, %Wannier90: A Tool for Obtaining Maximally-Localised Wannier Functions, 
Comput. Phys. Commun. 178, 685 (2008).

\bibitem{hse} J. Heyd, G. E. Scuseria, and M. Ernzerhof, %Hybrid functionals based on a screened Coulomb potential, 
J. Chem. Phys. 118, 8207 (2003); 124, 219906(E) (2006).

\bibitem{mp} J. Monkhorst and J. D. Pack, %Special points for Brillouin-zone integrations. 
Phys. Rev. B {\bf 13}, 5188 (1976).

\bibitem{sm} See Supplemental Material at [URL will be inserted by publisher] for additional details on structural, electronic, ferroelectric properties, on the Rashba effect and on the evaluation of $Z_2$ topological invariants.

\bibitem{unpub} In the case of FE GeTe, structurally very close to SnTe, the
polarization has been suggested, both theoretically using first-principle calculations and 
experimentally via PFM measurements, to be intrinsically perpendicular to 
the GeTe(111) thin films' surfaces. (V. L. Deringer, M. Lumeij, and R. Dronskowski, J. Phys. Chem. {\bf 116}, 15801 (2012); C. Rinaldi et al., to be published)

\bibitem{superf_calcs}
 B. M. Wojek, R. Buczko, S. Safaei, P. Dziawa, B. J. Kowalski, M. H. Berntsen, T. Balasubramanian,
 M. Leandersson, A. Szczerbakow, P. Kacman, T. Story, and O. Tjernberg, Phys. Rev. {\bf B} 87, 115106 (2013); 
 C. M. Polley, P. Dziawa, A. Reszka, A. Szczerbakow, R. Minikayev, J. Z. Domagala, S. Safaei, P. Kacman, R. Buczko, J. Adell, M. H. Berntsen, B. M. Wojek, O. Tjernberg, B. J. Kowalski, T. Story, and T. Balasubramanian, Phys. Rev. {\bf B} 89, 075317 (2014);
 J. Liu, W. Duan, and L. Fu, Phys. Rev. {\bf B} 88, 241303 (2013).

\bibitem{bahramy} M. S. Bahramy , R. Arita , N. Nagaosa , Phys. Rev. B {\bf 84},
041202(R) (2011).

\bibitem{fu} J. Liu, W. Duan and L. Fu, Phys. Rev. B {\bf 88}, 241303(R) (2013).

\bibitem{buco} S. Safaei, P. Kacman and R. Buczko, Phys. Rev. B {\bf 88}, 045305 (2013).

\bibitem{yongwang} Y. Wang, X. Liu, J. D. Burton, S. S. Jaswal, and E. Y. Tsymbal, Phys. Rev. Lett. {\bf 109}, 247601 (2012).

\bibitem{gesnte} A.I. Lebedev, I.A. Sluchinskaya, V.N. Demin and I.H. Munro JETP Lett. {\bf 63},
   635 (1996); A.I. Lebedev, I.A. Sluchinskaya, V.N. Demin and I.H. Munro Phase Transitions {\bf 60}, 67 (1997).

\end{thebibliography}
\end{document}